\documentclass[twocolumn, 11pt]{article}
\usepackage{array, float}
\usepackage[left=18mm, right=18mm, top=26mm, bottom=24mm]{geometry}
\usepackage{newtxtext}
\usepackage{amsmath, amssymb}
\usepackage{mathrsfs}
\usepackage{tabularx}
\usepackage{graphicx}
\usepackage{dcolumn}
\usepackage{setspace}
\usepackage{pgfpages}
\makeatletter
\usepackage{dcolumn}
\usepackage{url}
\usepackage{comment}
\usepackage{lineno}
\usepackage{cite}
\usepackage[numbers,sort&compress]{natbib}

\usepackage{abstract}

\usepackage{setspace}
\usepackage[font=small, format=plain, labelfont=bf, up, textfont=normal, up, justification=justified,singlelinecheck=false]{caption}

\usepackage[affil-it]{authblk}

\let\OLDthebibliography\thebibliography
\renewcommand\thebibliography[1]{
\OLDthebibliography{#1}
\setlength{\parskip}{0pt}
\setlength{\itemsep}{0pt plus 0.3ex}
}

\title{Hinderance of cooperation by individual solutions: Evolutionary dynamics of three-strategy games combining the prisoner's dilemma and stag hunt}
\author{Hirofumi Takesue\thanks{Electronic address: \texttt{hir.takesue@gmail.com}}}
\affil{Tokyo Metropolitan University}
\date{}

\begin{document}

\twocolumn[

\maketitle

\begin{onecolabstract}
We considered a three-strategy game with the characteristics of the prisoner's dilemma and stag hunt games. This game was inspired by recent experimental studies that elucidated the role of individual solutions. People who adopt individual solutions do not free-ride on the cooperative efforts of others, but instead attempt to solve the problem only to the extent necessary to prevent an impact on themselves. We argue that individual solutions play a role similar to that of defection in the stag hunt game, and this study examined the effects of orthodox free-riding and the individual solution on the evolution of cooperation. Our analysis revealed that a state in which all of the agents adopt the individual solution is the only stable equilibrium in the well-mixed population. Interactions on a square lattice led to modest improvements in cooperation levels, which were mainly sustained by cyclic dominance. Payoff values favorable to cooperation resulted in full cooperation, but rare mutations could hamper the cooperative equilibrium and support the dominance of the individual solution. Our analysis echoed experimental observations and illustrated that overcoming the reliance on individual solutions is critical in understanding the evolution of cooperation.
\\\\
\end{onecolabstract}
]
\saythanks

\section*{Introduction}
Cooperation can evolve through various routes \citep{Nowak2006}. Canonical models of social dilemmas predict that cooperation easily dissipates because free-riders can enjoy the collective benefits without bearing the cost of cooperation. However, cooperation is widely observed in animal and human societies, and research in various fields has proposed mechanisms that support the evolution of cooperation. Knowledge on notable mechanisms such as kin selection \citep{West2007}, direct reciprocity \citep{Rossetti2024}, indirect reciprocity \citep{Nowak2005, Xia2023}, spatial reciprocity \citep{Perc2017}, partner choice \citep{Roberts2021}, and punishment \citep{Fehr2018} has accumulated. These mechanisms offer routes of the evolution of cooperation.

Despite progress in understanding the evolutionary roots of cooperation, recent human experiments indicated that individual solutions represent another challenge to the emergence of cooperation \citep{Gross2019, Gross2020}. In social dilemmas, both free-riding and individual solutions are viable options for individuals who do not contribute to collective solutions. People who rely on individual solutions attempt to solve problems only to the extent necessary to prevent harm to themselves opposed to free-riding on the cooperative effort of others. For instance, people might avoid the detrimental effects of climate change by moving to a less susceptible place \citep{Gross2020}. This individual solution resolves the issues of specific individuals without contributing to collective benefits, i.e., solving environmental problems. Individual solutions are viable in other areas such as social security.  

In pioneering experiments, participants had three behavioral options: investment in a public pool, investment in an individual pool, and keeping private resources \citep{Gross2019, Gross2020}. These options represent cooperation, individual solutions, and free-riding, respectively. If the total investment in the public pool reaches the threshold, all group members can enjoy the benefits of the collective solution. Free-riders can acquire the largest payoffs by keeping their own resources and enjoying the benefits of public investment provided by others. Conversely, they gain nothing from the failure of the provision of collective benefits, which constitutes the orthodox social dilemma. Investment in the individual pool can ensure some payoffs regardless of the success of the collective solution. However, if each individual adopts this individual solution, the outcome will be less efficient than collective cooperation because of the lack of scale merit. The experiments revealed that many participants selected the individual solution, often generating less efficient social outcomes.  

Motivated by this observation, we analyzed the evolutionary dynamics of the stylized version of the experimental game. Although the original game involved complex factors, such as repeated group interactions, we considered a simple two-person, three-strategy game that partially captured the essence of the experiment. The two options, namely investment in the public pool and keeping resources, correspond to cooperation and defection, respectively, in the prisoner's dilemma game (PDG). Another action, i.e., investment in the individual pool, is similar to defection in the stag hunt game (SHG). Although investing in the individual pool has a less efficient outcome than collective cooperation, it ensures some payoffs, and it is better than solitary investment in the collective pool. This is analogous to defection (i.e., hunting a hare) in the SHG. Although the benefit of hunting a hare is smaller than that of capturing a stag, the action ensures certain benefits without the risk of hunting a stag solitarily. We therefore analyzed a three-strategy game capturing characteristics of both the PDG and SHG. We called this game PDG with an individual solution following the pioneering experimental studies. A previous study examined the combination of the snowdrift game and SHG, namely the snowdrift game with an individual solution \citep{Takesue2025}. 

Our analysis of the evolutionary dynamics of the PDG with an individual solution demonstrated the dominance of the individual solution. Straightforward analysis of a well-mixed population model illustrated that agents adopting the individual solution occupy the population at the unique asymptotically stable rest point of the replicator dynamics. In addition, we examined the interactions in networked populations, which is an intensively studied mechanism that supports cooperation \citep{Perc2017}. The Monte Carlo simulations indicated that networks support the evolution of cooperation to some extent, but the individual solution remained relatively dominant. Payoff values conducive to cooperation often led to full cooperation. However, the introduction of small mutations hampered the cooperative equilibrium, and the individual solution dominated the population. These results are subsequently discussed in comparison to similar games, including the voluntary PDG \citep{Szabo2002} and PDG with an exit option \citep{Shen2021}. Our study demonstrates that reliance on the individual solution can impede the evolution of cooperation, in line with the observations of previous experiments.

\section*{Model}
We considered a three-strategy game combining the PDG and SHG based on prior research \citep{Gross2019}. The three strategies in the PDG with an individual solution were cooperation ($C$), defection ($D$), and the individual solution ($I$). We considered the following payoff matrix:
\[
\begin{array}{c}
C\\ D\\ I\\
\end{array}
\begin{pmatrix}
b-c & b-2c & b-2c \\
b   & 0 & 0 \\
b-c_I & b-c_I & b-c_I \\
\end{pmatrix}.
\]
Cooperation generates benefits $b$ for \textit{both} agents. The cost to generate the benefit is $c$ when mutual cooperation is achieved. For simplicity, we assumed that cooperation costs twice as much when the partner does not cooperate because the agent individually covers the cost to produce the benefit. Defectors can enjoy the benefit with no cost when interacting with a cooperator, but they gain nothing when the partner does not provide collective benefits. The individual solution ensures the same payoff of $b-c_I (> 0)$ regardless of the partner's action. The collective benefits provided by cooperators do not affect payoffs because the individual solution already achieves the required goal. The individual solution is more costly than mutual cooperation, but it costs less than providing collective benefits alone. We therefore assumed that $c < c_I < 2c$. Under the assumption that $c < b < 2c$, this game is reduced to the PDG when the actions $C$ and $D$ are considered. Conversely, it is reduced to the SHG when the actions $C$ and $I$ are considered. In summary, we assumed that $c < c_I < b < 2c$. In addition, we examined the case in which a different payoff matrix that can take flexible payoff values. This payoff matrix will be introduced in the Results section.

We examined the evolutionary dynamics of this game in canonical settings \citep{Sigmund2010, Perc2017}. First, we briefly examine the case of a well-mixed population in which the system evolves under replicator dynamics. Second, we examine the case of a structured population where $L^2$ agents are located on a square lattice with periodic boundary conditions. Starting with a random strategy configuration, one randomly selected agent $i$ can imitate the strategy of a randomly selected neighbor $j$ in each elementary step of the Monte Carlo simulation. The imitation probability is $1 / [1 + \exp(\beta (\Pi_i - \Pi_j))]$, where $\Pi_i$ and $\Pi_j$ are payoffs of each agent accumulated over games with four neighbors. The intensity of the selection is controlled by the temperature-like parameter ($\beta$). This rule assures that strategies that achieve large payoffs in the local environment tend to proliferate by imitation. In addition, mutations can occur with a probability of $\mu$, and agent $i$ adopts a randomly selected strategy. Small $\beta$ and large $\mu$ introduce noise into the evolutionary process. During a Monte Carlo step (MCS), agents update their strategy once on average.

Our quantities of interest were the frequencies of cooperation, defection, and the individual solution, which were denoted as $\rho_C$, $\rho_D$, and $\rho_I$, respectively. For the well-mixed population, we investigated these values by examining the asymptotically stable rest points. For the structured population, we calculated the strategy frequencies in the stationary states in the Monte Carlo simulations. Simulations typically consisted of the relaxation process that lasts $2 \times 10^5$ MCSs and the sampling process that lasts $2 \times 10^4$ MCSs. We calculated the average values of the four simulation runs. 

\section*{Results}
A straightforward analysis of the payoff matrix revealed that agents who adopt $I$ occupy the well-mixed population. States in which all the agents adopted $C$ were unstable because of the invasion by $D$ as in the standard PDG. States in which all agents adopted $D$ were not stable because agents who adopt the individual solution gain a larger payoff than defectors when the partner adopts defection ($b-c_I > 0$). There is a mixed strategy Nash equilibrium that is occupied by agents who adopt $C$ and $I$. This equilibrium was unstable similarly as noted in the two-strategy SHG. A state in which all agents adopted $I$ was the only asymptotically stable rest point. The stability was confirmed by the fact that selecting $I$ realizes the largest payoff when the partner also adopts $I$ ($b-c_I > b-2c$ and $b-c_I > 0$). This simple analysis demonstrated the advantage of the individual solution. 

\begin{figure}[tbp]
\centering
\vspace{5mm}
\includegraphics[width = 85mm, trim= 0 0 0 0]{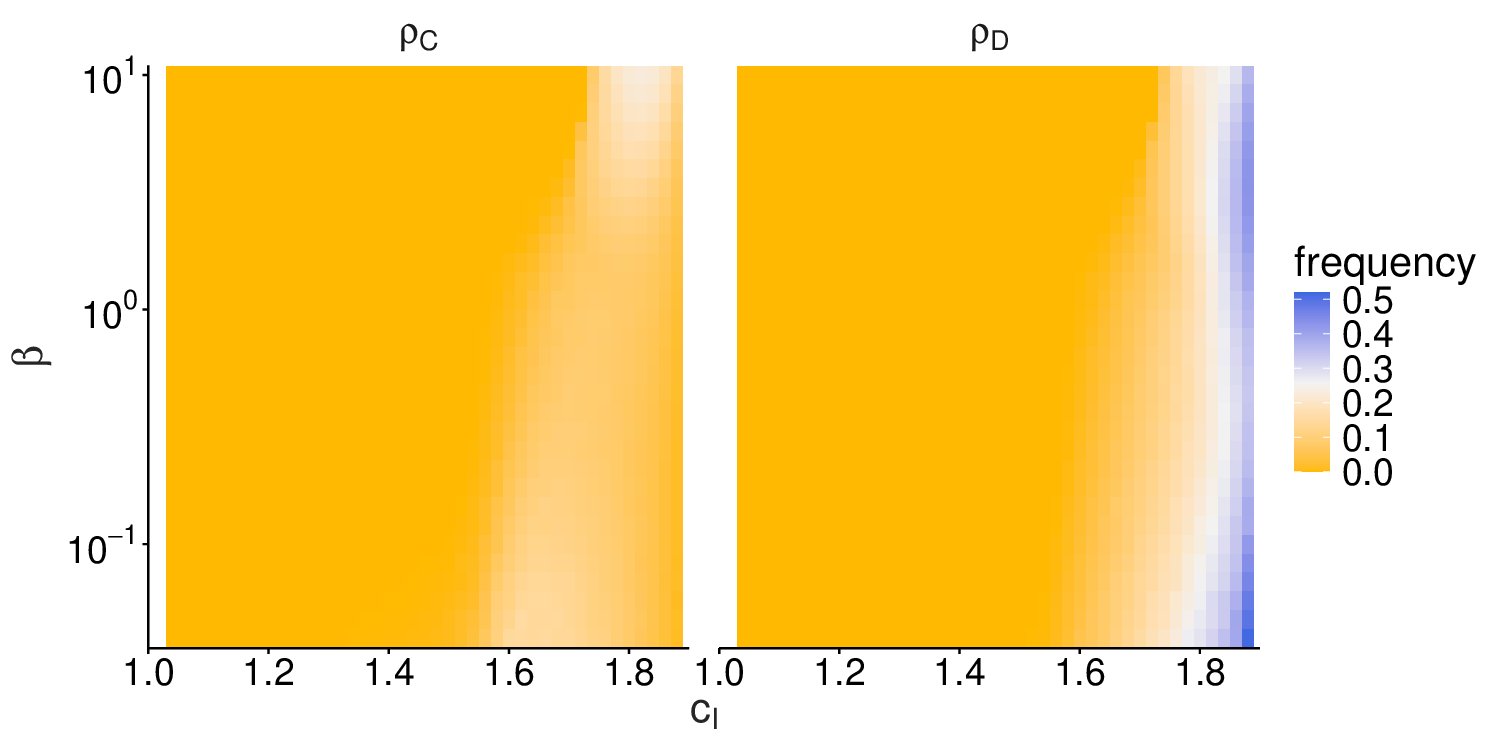}
\caption{\small Modest cooperation enhancement in networks with $b = 1.9$. Strategy frequencies on a lattice are reported as a function of the cost of the individual solution and the selection intensity. The individual solution dominates the population when its cost is small as the values of $\rho_C$ and $\rho_D$ are almost zero. The cooperation frequency peaks with the moderate cost of the individual solution. Other parameters: $L = 120, c = 1$, and $\mu = 10^{-4}$.}
\label{rho_c_beta}
\end{figure}

The individual solution also achieved large frequencies in a networked population, although spatial competition leads to complexity. Figure~\ref{rho_c_beta} reports the values of $\rho_C$ and $\rho_D$ when $b = 1.9$. Similar to the case in the well-mixed population, the individual solution was predominant when the cost of the individual solution ($c_I$) was sufficiently small, whereas $\rho_C$ and $\rho_D$ were negligible. Cooperation survived when $c_I$ became large. $\rho_C$ peaked with intermediate values of $c_I$, as documented in the left panel. Cooperation enhancement in the structured population was modest, as the maximum value of $\rho_C$ was 0.23.

\begin{figure}[tbp]
\centering
\vspace{5mm}
\includegraphics[width = 85mm, trim= 0 0 0 0]{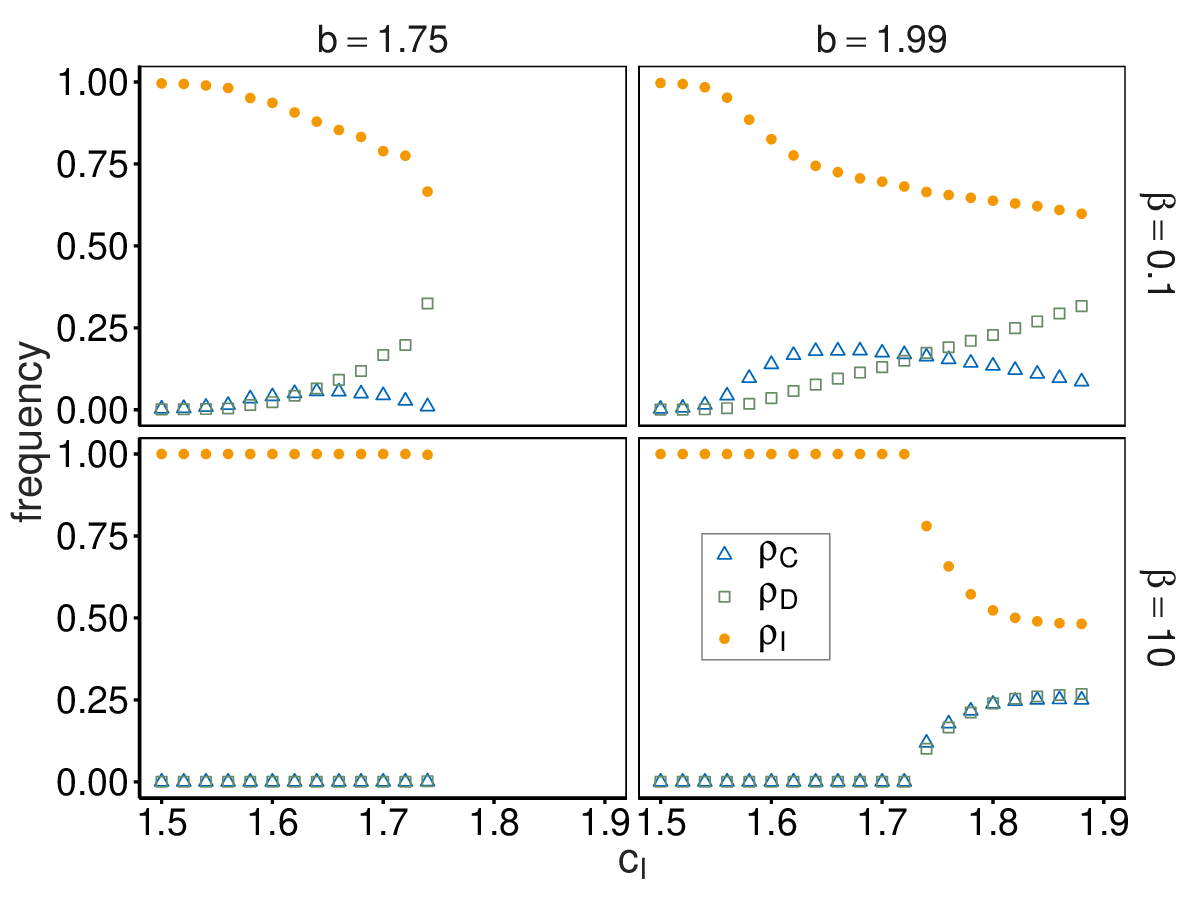}
\caption{\small Dominance of the individual solution and coexistence of the three strategies in the networks. Strategy frequencies on a lattice are reported. The basic patterns in the Figure~\ref{rho_c_beta} are replicated with $b = 1.75$ and $b = 1.99$. Other parameters: $L = 120, c = 1$, and $\mu = 10^{-4}$.}
\label{rho_c_b_beta}
\end{figure}

Figure~\ref{rho_c_b_beta} reports the strategy frequencies for different values of $b$ (1.75 and 1.99), and similar patterns were observed. The small cost of the individual solution ($c_I$) led to an equilibrium in which almost all agents adopted the individual solution. $\rho_C$ and $\rho_D$ gradually increased as the cost of the individual solution increased, and the three strategies coexisted.

\begin{figure}[tbp]
\centering
\vspace{5mm}
\includegraphics[width = 80mm, trim= 0 0 0 0]{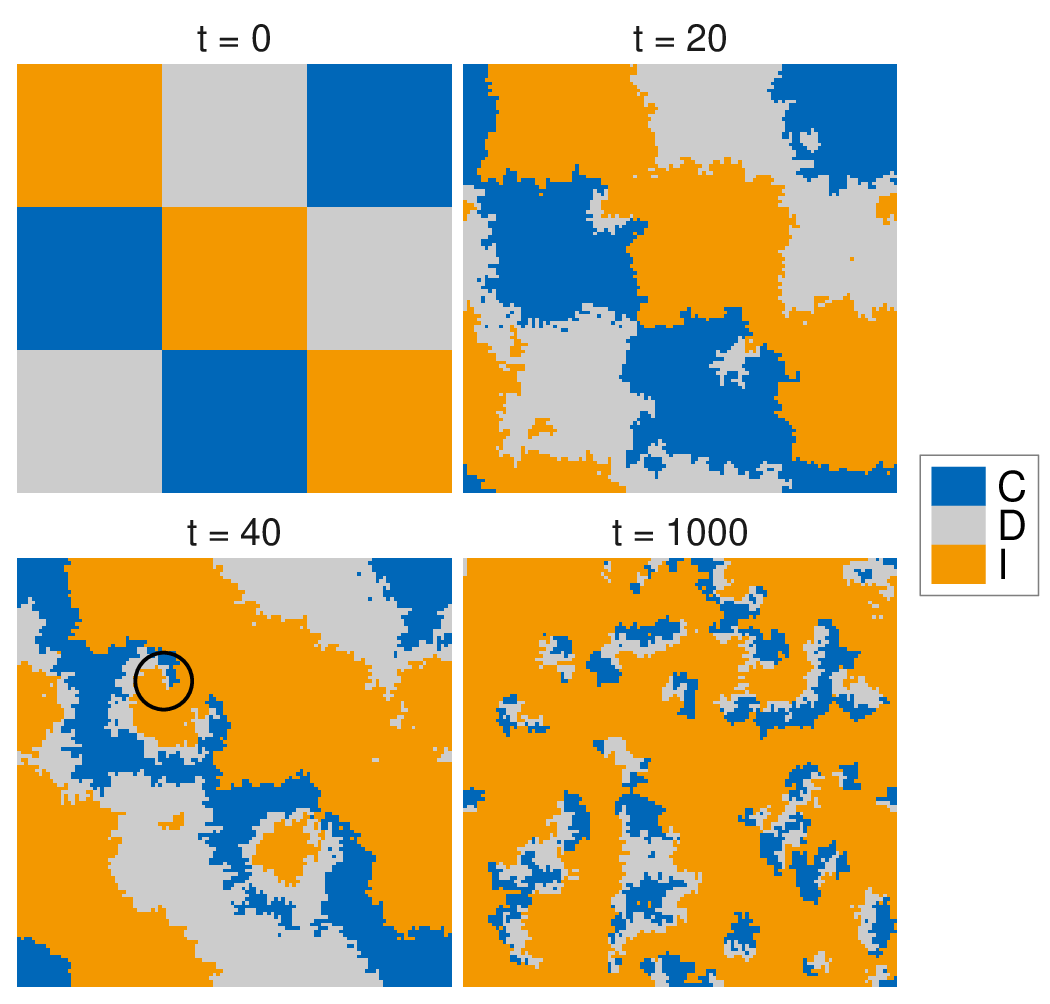}
\caption{\small Emergence of cyclic dominance on networks. An example of the evolutionary trajectory on a lattice is reported. Cyclic dominance, in which defection dominates cooperation, cooperation dominates the individual solution, and the individual solution dominates defection, is observed. A black circle encircles an example of the rotating spirals observed with cyclic dominance. Parameters: $L = 120, b = 1.95, c = 1, c_I = 1.76, \beta = 10$, and $\mu = 10^{-4}$.}
\label{lattice_time}
\end{figure}

This coexistence of the three strategies was achieved through spatial cyclic dominance. Figure~\ref{lattice_time} presents the time evolution of the arrangement of the three strategies on the lattice to visualize evolution of the system. To clarify the spatial pattern of system evolution, the simulation started with the prepared initial state ($t = 0$; see Ref.~\citep{Szolnoki2011b} for a seminal study that adopted this technique). As the simulation proceeded, the patterns of spatial competition emerged. Agents employing $C$ adopted $D$; those employing $D$ adopted $I$; those employing $I$ adopted $C$. The figure presents transitions between strategies moving in the upper right direction ($t = 20$ and $t = 40$). In addition, rotating spirals that typically accompany cyclic dominance \citep{Szolnoki2005} are marked by a black circle. After some relaxation processes, clusters of cooperators and defectors are surrounded by the agents who adopt the individual solution ($t = 1 000$).

In addition, we considered a payoff matrix that takes flexible payoff values and conducted additional simulations. The considered matrix is
\[
\begin{array}{c}
C\\ D\\ I\\
\end{array}
\begin{pmatrix}
1 & S & S \\
T & 0 & 0 \\
\sigma & \sigma & \sigma \\
\end{pmatrix}.
\]
The same payoff order was preserved if we assumed that $S < 0 < \sigma < 1 < T$. Because a common parameter does not appear in the payoff of multiple strategies, this matrix permitted flexible analysis. This flexibility has certain value, although the payoff matrix deviated from the motivating setting of this study. This game is equivalent to the PDG with an exit option \citep{Shen2021} when the value of $S$ is zero. In other words, we adopted the PDG, whereas the previous study adopted a weak PDG \citep{Nowak1992}. Therefore, we refer to observations in the PDG with an exit option for the purpose of comparison. Because the PDG with an exit option was examined without mutation, we often compared the cases of $\mu = 0$ and $\mu > 0$.

\begin{figure*}[tbp]
\centering
\vspace{5mm}
\includegraphics[width = 130mm, trim= 0 0 0 0]{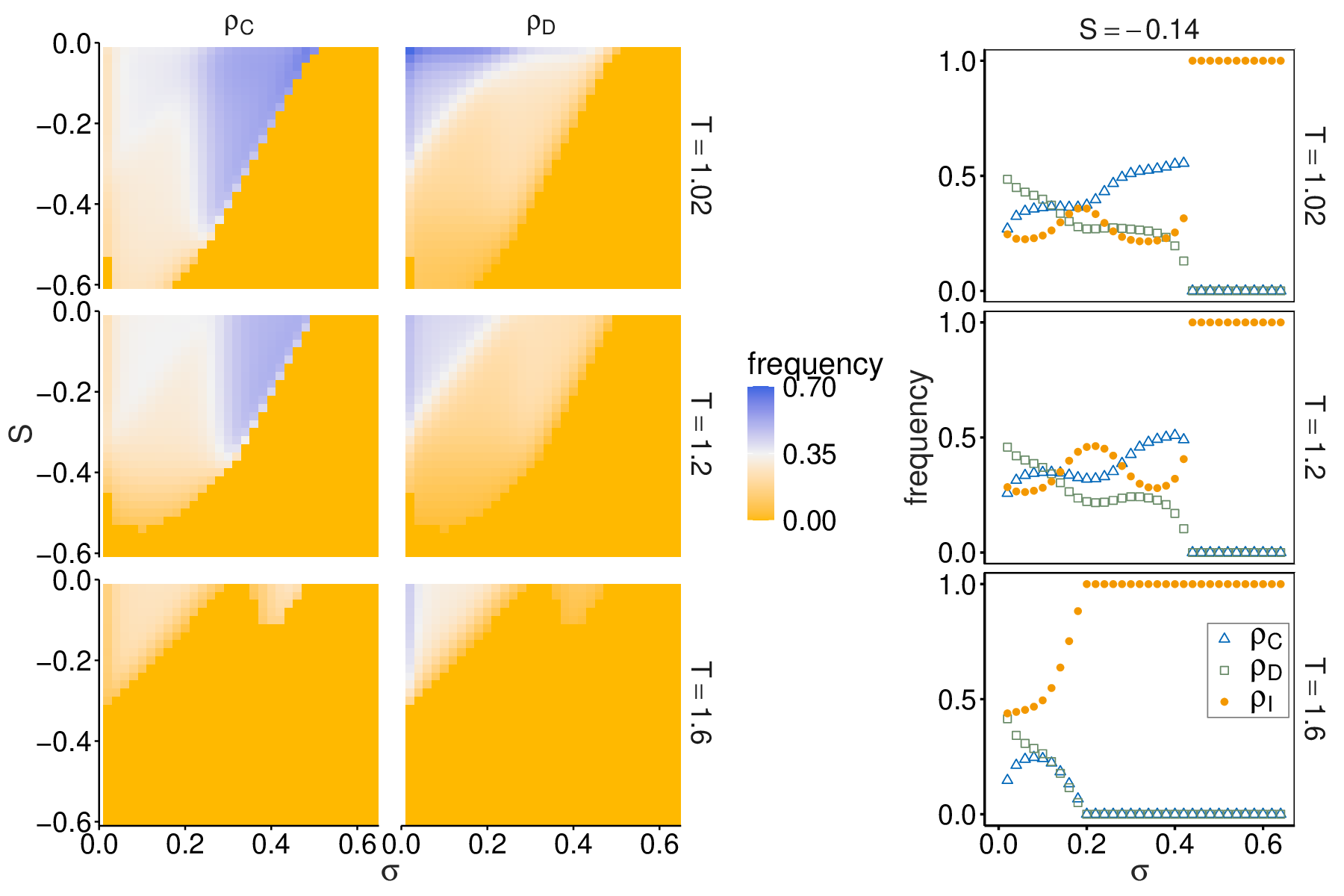}
\caption{\small Dominance of the individual solution and coexistence of the three strategies with the different payoff matrix. Strategy frequencies with the different payoff matrix is reported. The dominance of the individual solution appears with large $T$, small $S$, and large $\sigma$, whereas the coexistence of the three strategies appears with small $T$, large $S$, and small $\sigma$ (left heatmaps). Sharp transitions between the different strategy compositions are often observed (right panels). The values of $L$ ranged from 120 to 480. Other parameters: $\beta = 10$ and $\mu = 10^{-4}$.}
\label{rho_STS}
\end{figure*}

A qualitatively similar pattern was observed with this setting when $\mu = 10^{-4}$ (Figure~\ref{rho_STS}). The observed outcomes were the dominance of the individual solution or the coexistence of the three strategies. The small absolute size of the sucker's payoff ($S$), the small temptation to defect ($T$), and the small payoffs for the individual solution ($\sigma$) basically fostered the coexistence of the three strategies, as indicated by the left heatmaps that present $\rho_C$ and $\rho_D$. As expected, the cooperation levels were enhanced by the small absolute size of $S$ and small $T$. The maximum cooperation levels were observed with moderate values of $\sigma$, similar to the effect of $c_I$ (Figure~\ref{rho_c_beta}). Cooperators often become the majority in these advantageous situations. Nonetheless, the heatmaps also demonstrate that the dominance of the individual solution is assured when its payoff is approximately half of that for mutual cooperation (0.5), as indicated by the negligible values of $\rho_C$ and $\rho_D$. The main difference from the original setting was the transitions between the dominance of the individual solution and the coexistence of the three strategies. The original setting resulted in continuous transitions (Figure~\ref{rho_c_b_beta}), whereas a sharp increase (decrease) in the value of $\rho_I$ ($\rho_C$) was observed when $T = 1.02$ and $T = 1.2$ (the right three plots in Figure~\ref{rho_STS}). Smooth transitions were observed in the harsh environment for the evolution of cooperation ($T = 1.6$).

\begin{figure}[tbp]
\centering
\vspace{5mm}
\includegraphics[width = 80mm, trim= 0 0 0 0]{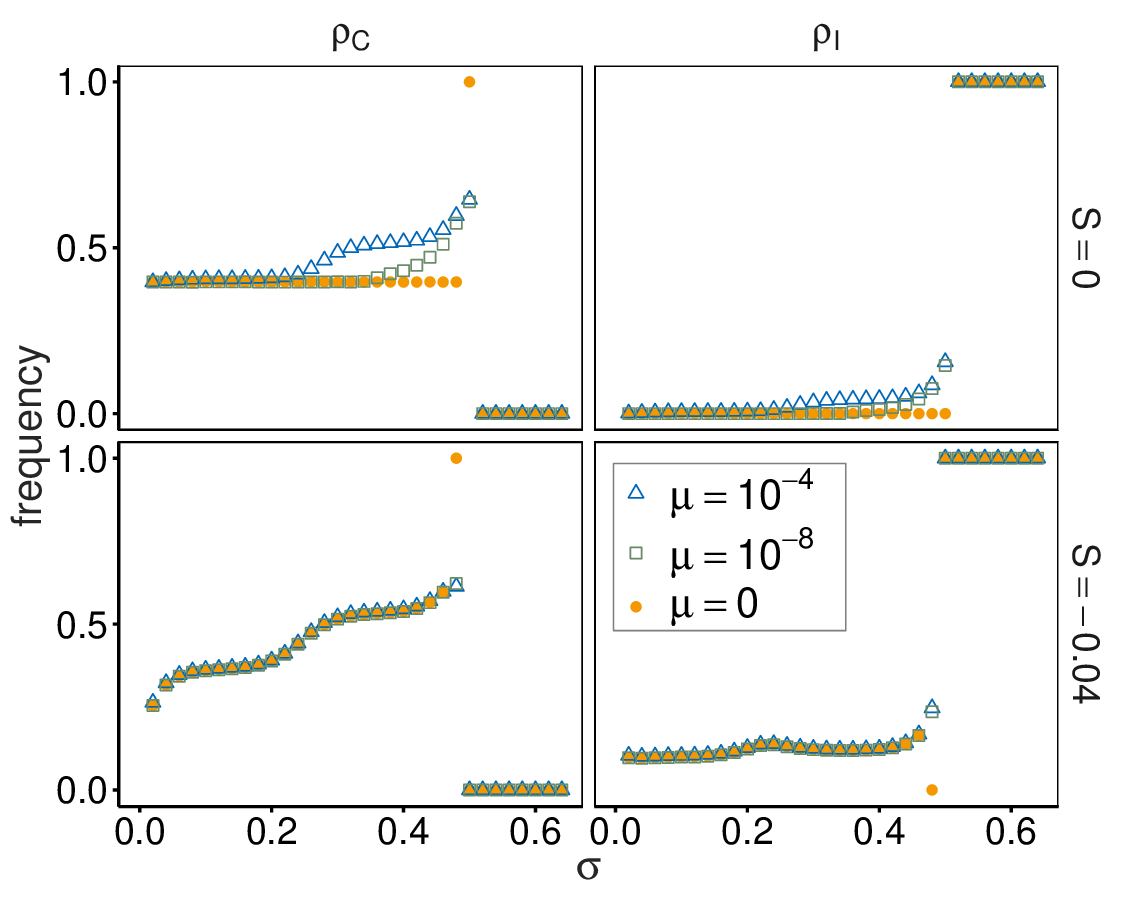}
\caption{\small Disappearance of \textit{C}--\textit{D} equilibrium by the negative sucker's payoff ($S$). Strategy frequencies at $T = 1.02$ are reported. The phase in which cooperation and defection occupy the population is observed at $S = 0$, but it is removed when $S = -0.04$. Full cooperation is fragile to the introduction of the mutation. The values of $L$ range from 120 to 720. The value of $\beta$ was 10.}
\label{rho_STS_T102}
\end{figure}

To understand the behavior of the system, we focused on the effects of $S$, $\sigma$, and $\mu$. The results at $T = 1.02$ illustrated that negative $S$ erases the coexistence of cooperation and defection (Figure~\ref{rho_STS_T102}). At $S = 0$, we observed the phase in which the population was occupied by the mixture of cooperation and defection, and the frequencies of the individual solution in the right panels were zero when $\mu = 0$. A previous study observed this two-strategy solution with small $T$ \citep{Shen2021}. This phase was observed if the payoffs for the individual solution were sufficiently small ($\sigma \leq 0.5$). This phase disappeared at $S = -0.04$, being replaced by the coexistence of the three strategies. In other words, the phase observed with the combination of \textit{weak} PDG and SHG is vulnerable to negative $S$.

\begin{figure}[tbp]
\centering
\vspace{5mm}
\includegraphics[width = 80mm, trim= 0 0 0 0]{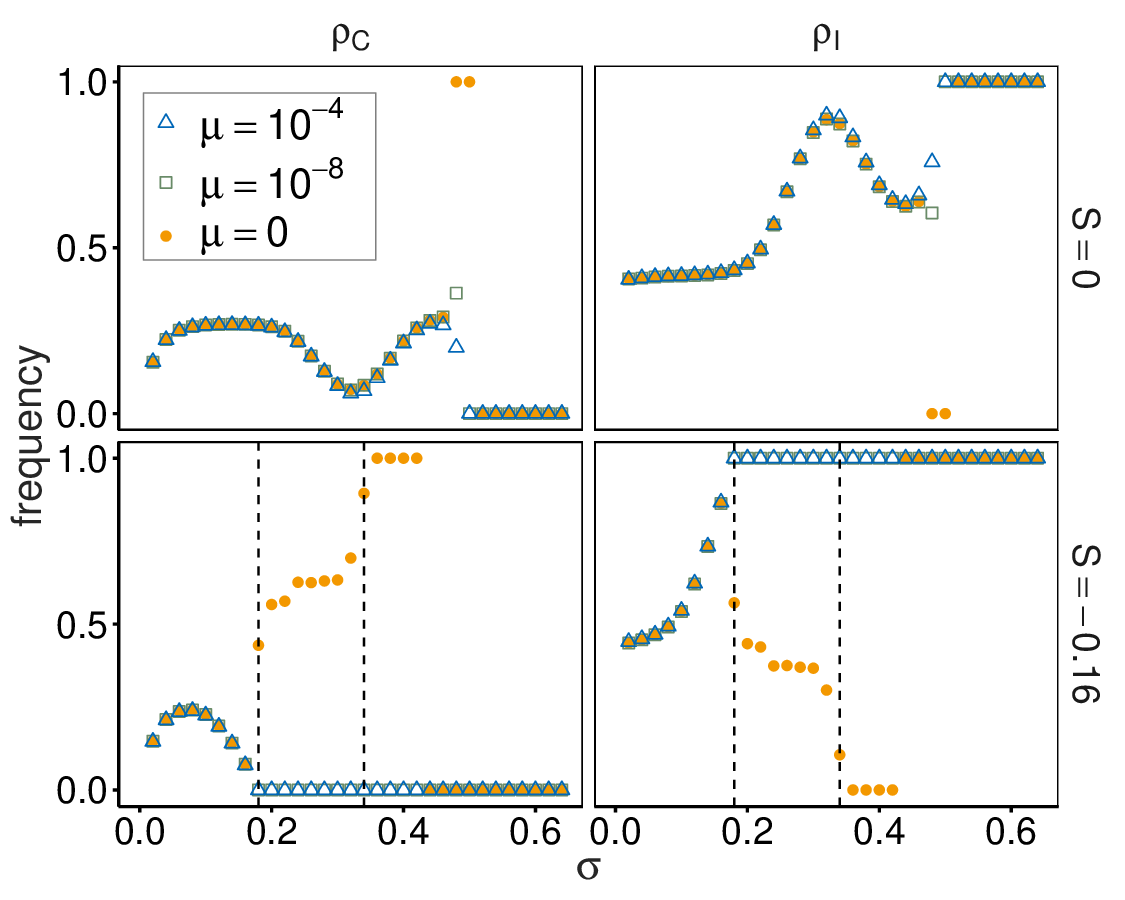}
\caption{\small Expansion of the cooperative equilibrium by the negative sucker's payoff ($S$) and its fragility to mutation. Strategy frequencies at $T = 1.6$ are reported. Full cooperation is more likely when $S = -0.16$. The parameter region in which bistability occurred is denoted by the dashed lines. In this region, we conducted 1 000 simulation runs with $L =$ 1 000 and calculated the average frequencies. Full cooperation was fragile to the introduction of the mutation. The values of $L$ ranged from 120 to 1 000. The value of $\beta$ was10.}
\label{rho_STS_T160}
\end{figure}

Under different conditions ($T = 1.6$), a negative value of $S$ often supported cooperation when $\mu = 0$ (Figure~\ref{rho_STS_T160}). When $S = 0$, the frequency of cooperation fluctuated between 0.07 and 0.29, and full cooperation suddenly appeared when $\sigma$ was 0.46. Full cooperation was observed over a limited range of parameter values. When $S = -0.16$, full cooperation appeared with a $\sigma$ value of 0.18, and bistability characterized the system. This bistability was observed with a relatively large system size ($L =$ 1 000). The population was dominated by cooperation or individual solutions, and the probability that the system reaches full cooperation increased as the payoff for the individual solution increased. This bistability was derived from the nature of the SHG. At $\sigma = 0.38$, the system reached full cooperation in all simulation runs. Consequently, a comparison of the results at $S = 0$ and $S = -0.16$ illustrated that full cooperation occurred over a wider range of parameter values when $S = -0.16$.

Mutation favors individual solutions, and the full cooperation observed in the analysis was highly fragile to mutation. Full cooperation (including that in bistability) is presented in Figures~\ref{rho_STS_T102} and \ref{rho_STS_T160}. However, cooperation significantly decreased in frequency once the mutation was introduced, and the individual solution emerged. In many cases, the frequency of cooperation declined from one to zero. The probability of mutation does not need to be large to eliminate cooperation, as the values of $\mu$ in the figures are $10^{-8}$ and $10^{-4}$. This suggests that predominance of cooperation requires a noiseless environment. Because of this sensitivity to mutation, the enhancement of cooperation attributable by the sucker's payoff ($S$) disappears, and the cooperation frequency monotonically decreases as the absolute size of $S$ increases in cases with mutation (Figure~\ref{rho_STS}).

\begin{figure}[tbp]
\centering
\vspace{5mm}
\includegraphics[width = 80mm, trim= 0 0 0 0]{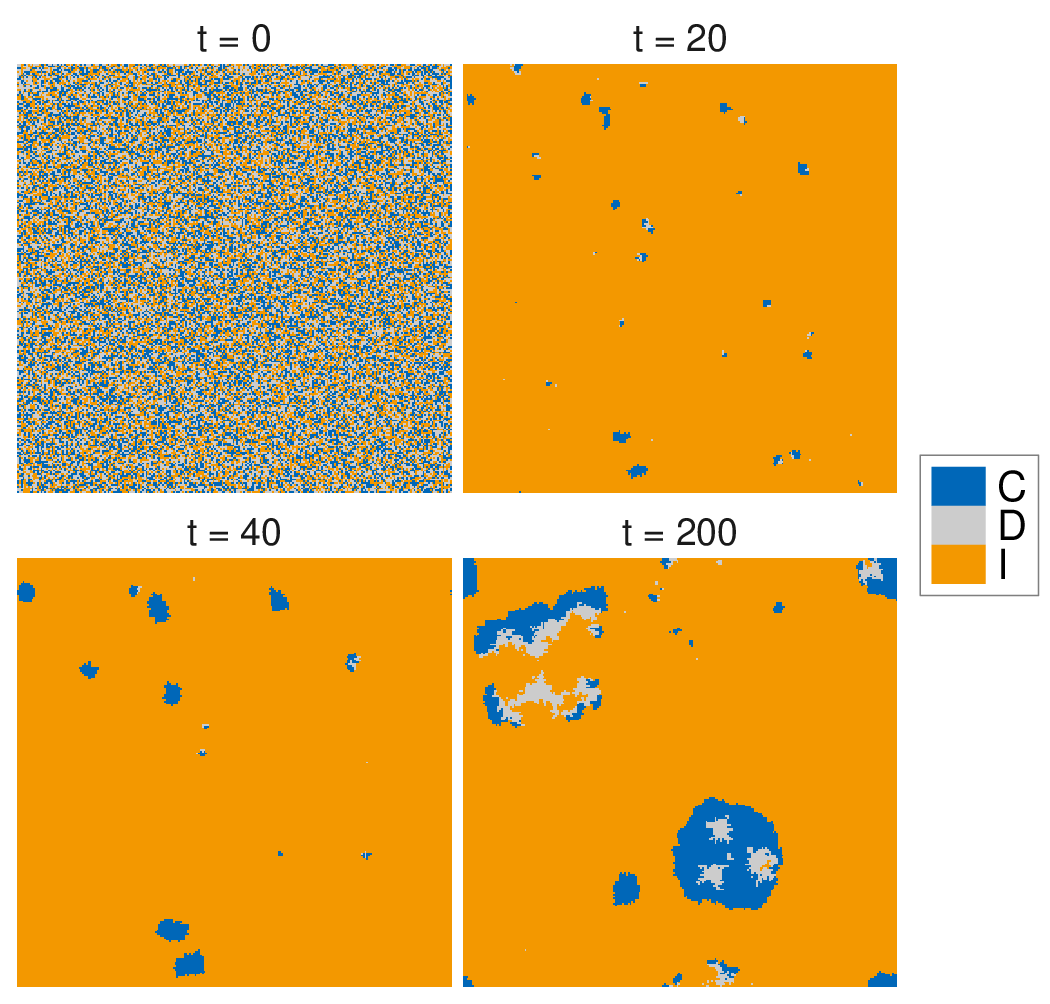}
\caption{\small Cooperation clusters broken by mutation. An example of the evolutionary trajectory on a lattice is presented. Defection introduced by mutation infiltrates cooperative clusters, which leads to the dominance of the individual solution. Parameters: $L = 240, S = -0.16, T = 1.6, \sigma = 0.38, \beta = 10$, and $\mu = 10^{-4}$.}
\label{lattice_time_STS}
\end{figure}

The prohibition of cooperation by mutation can be visualized by the time evolution of the strategies on the lattice. Figure~\ref{lattice_time_STS} presents spatial evolution of the three strategies. From the random initial configuration ($t = 0$), the evolutionary dynamics led to the formation of compact clusters of cooperators ($t = 20$ and 40). In the absence of mutation, these clusters of cooperators ultimately dominated the entire population through the classic mechanism of spatial reciprocity (although the competitor is the individual solution in this case). Defection was suppressed by the individual solution when $\sigma$ was sufficiently large, permitting the expansion of cooperation through spatial reciprocity. However, mutations enabled defectors to infiltrate clusters of cooperators, and clusters of cooperators accompanied neighboring agents who adopt the defection strategy because of rare mutations ($t = 200$), thereby preventing the expansion of cooperation. As the individual solution produced larger payoffs than defection, mutation ultimately led to the proliferation of the individual solution.

\section*{Discussion}
We investigated the evolutionary dynamics of the PDG with an individual solution, which is a three-strategy game featuring the properties of the PDG and SHG. Considering only cooperation and defection reduces this game to a PDG, whereas considering only cooperation and the individual solution reduces the game to an SHG. The game was inspired by experimental studies demonstrating the significance of the individual solution \citep{Gross2019, Gross2020}. Our study found that individual solutions prevail in both well-mixed and structured populations. Individual solutions completely dominated the population under replicator dynamics. In a structured population, roc--paper--scissor dynamics supports cooperation to some extent, but the individual solution still achieves large frequencies in a wide range of parameter regions. The analysis adopting a flexible payoff matrix also indicated that the survival of cooperation is mostly maintained through the coexistence of the three strategies. These results were consistent with those of the motivating experimental studies in that people often resort to the individual solution.

The game considered in this study is similar to the widely studied voluntary PDG (and its extensions) \citep{Szabo2002, Wu2005, Guo2017, Lu2017, Deng2018, Wang2018a, Cardinot2019, Yamamoto2019, Guo2020c, Tao2020, Zhang2020a, Khatun2024}. The loner strategy introduced in this game permitted nonparticipation in game interactions \citep{Szabo2002}. This game had an interior equilibrium that included cooperation, defection, and the loner strategy. Conversely, the game in the present study only had one pure strategy equilibrium that was occupied by the individual solution. Despite this different property in well-mixed populations, structured populations had similar dynamics. In the voluntary PDG, cooperation invades the loner strategy, the loner strategy invades defection, and defection invades cooperation \citep{Szabo2002, Cardinot2019, Guo2020c, Tao2020, Jia2024}. In the PDG with an individual solution, similar cyclic dominance is replicated in networks when the loner strategy is replaced by the individual solution. Cyclic dominance plays a crucial role in the evolution of cooperation beyond the voluntary PDG \citep{Szolnoki2014c, Szolnoki2017a, Szolnoki2018, Chu2019, Chu2020, Hu2020, Szolnoki2020, Gao2021a, Podder2021a, Lee2022, Liu2022b, Chen2023, Shen2025}. 

The structure of the PDG with an individual solution is also similar to that of the PDG with an exit option \citep{Shen2021}. In the framework of this study, the PDG with an exit option combined the \textit{weak} PDG and SHG, and therefore, the sucker's payoff ($S$) was set to zero. Our study demonstrated that the sucker's payoff had equivocal effects on evolutionary dynamics. An increase in the absolute size of $S$ disrupted the equilibrium consisting of cooperation and defection and begat the coexistence of the three strategies. By contrast, a negative sucker's payoff fostered the emergence of a full cooperation equilibrium, and the individual solution disappeared. This is similar to the observation in the PDG with an exit option that the increase in temptation to defect ($T$) supports the dominance of cooperation. Full cooperation is often characterized by bistability, which was not observed in the PDG with an exit option. 

However, the most powerful factor was the introduction of mutation. Mutation disrupts the full cooperation equilibrium, and the individual solution often occupies the entire population. The collapse of the cooperation equilibrium is observed regardless of the type of the combined PDG. Previous studies analyzed the role of mutation \citep{Adami2016, Ichinose2018a}, and introduction of mutation alters strategy frequencies \citep{Helbing2009, Takesue2019a} and the speed of convergence \citep{Helbing2010d} in spatial evolutionary games. Our research revealed a new example of the influence of mutations on evolutionary games in networks. The PDG with an exit option displays rich nontrivial behavior including cyclic dominance, a local surge of cooperation levels, and the two-strategy solution of cooperation and defection \citep{Shen2021}. Consideration of the sucker's payoff and mutation in our analysis revealed that cyclic dominance was the most robust mechanism sustaining cooperation in the game investigated in this study. Conversely, some phases in the PDG with an exit option were also observed in a different game \citep{Takesue2025}. Examining when the nontrivial and rich behaviors observed in the previous study \citep{Shen2021} occur could be a meaningful task. 

The consideration of multiple games was not limited to this study, as extensive studies have been conducted under the framework of multi-games \citep{Wang2014, Qin2017, Deng2018, Guo2019, Huang2019a, Li2019a, Deng2021b, Li2021b, Wu2021}. In multi-games, the sucker's payoff is positive for some agents but negative for others. Consequently, agents participate in two different games: the prisoner's dilemma and the snowdrift game. These studies found that multi-games enhance cooperation levels. Although the SHG is not widely considered in the literature, multi-games that consider the stag hunt also enhanced cooperation \citep{Cheng2023}. Our study also considered multiple games, but it introduced two games in a different manner. We believe that considering multiple games by considering multiple strategies might also improve our understanding of the evolution of cooperation.

Finally, we will discuss the limitations and potential extensions of this study. First, the natural extension considers group interactions \citep{Perc2013a}. This study considered a simple two-person game, but more accurate replication of experimental tasks and provision of collective solutions in societies will require the consideration of public goods games. Second, other strategies should be considered. For example, punishment is a prominent strategy that supports the emergence of cooperation \citep{Fehr1999}, and its role (in combination with spatial structure) is an important topic in the literature \citep{Helbing2010c, Chen2015, Szolnoki2017a, Yang2018, Wang2019d, Flores2021, Sun2021, Hua2023a, Han2024, Lee2024}. A study examining a PDG with an exit option found that the exit option supports the emergence of altruistic punishment \citep{Shen2025}. The role of punishment was investigated in the motivating experiments \citep{Gross2019}, and the consideration of punishment represents a natural extension in combination with spatial structure and mutation. We believe that considering individual solutions is a simple extension, but it will contribute to understanding the evolution of cooperation and its overlooked difficulty.


\end{document}